# Symbol repetition in interstellar communications: methods and observations

*William J. Crilly Jr.*

Green Bank Observatory, West Virginia, USA

*Abstract*— Discoverable interstellar communication signals are expected to exhibit al least one signal characteristic clearly distinct from random noise. A hypothesis is proposed that radio telescope received signals may contain transmitted $\Delta t\ \Delta f$ opposite circular polarized pulse pairs, conveying a combination of information content and discovery methods, including symbol repetition. Hypothetical signals are experimentally measured using a 26 foot diameter radio telescope, a chosen matched filter receiver, and machine post processing system. Measurements are expected to present likelihoods explained by an Additive White Gaussian Noise model, augmented to reduce radio frequency interference. In addition, measurements are expected to present no significant differences across a population of Right Ascension ranges, during long duration experiments. The hypothesis and experimental methods described in this paper are based on multiple radio telescope $\Delta t\ \Delta f$ polarized pulse pair experiments previously reported. (ref. arXiv:2105.03727, arXiv:2106.10168). In the current work, a Right Ascension filter spans twenty-one 0.3 hour Right Ascension bins over a 0 to 6.3 hr range, during a 143 day experiment. Apparent symbol repetition is measured and analyzed. The 5.25 ± 0.15 hr Right Ascension, -7.6° ± 1° Declination celestial direction has been associated with anomalous observations in previous work, and continues to present anomalies, having unknown cause.

## I. Introduction

Hypothetical discoverable interstellar communication signals may be guessed to exhibit certain characteristics. The signals may utilize energy-efficient information transfer methods, produce relatively low interference to other communication systems, and contain at least one signal design element that aids discovery. The signal discovery mechanism is expected to provide unique transmitter identification, and to contain one or more seeds for a receiver to learn the information in the signal. The design space of such signals is large, yet may be reduced by studying communication systems based on principles of mathematics and physics [1][2], assumed to be similarly understood at the transmitter and receiver. The analyzed existence of an interstellar coherence hole (ICH) communication method is an example of such similar shared principles.[2] The characteristics of signals and systems proposed in this paper are based on these principles.

In previous work, $\Delta t\ \Delta f$ opposite-polarized pulse pairs were examined as possible interstellar communication signals. [3][4] The arrival time and RF frequency differences between the polarized pulses in each pulse pair, i.e. $\Delta t\ \Delta f$ measured values, of decreasingly-sorted Signal to Noise Ratios (SNR) of the pulses were used to measure the pulse pairs' likelihood due to an Additive White Gaussian Noise (AWGN) model, augmented with robust RFI amelioration.

The augmented AWGN model was apparently refuted as a likely cause of the anomalies observed, in past work. Similar anomalies appear in the current work, and new anomalies that indicate a symbol repetition scenario have been observed. Experimental results of recent work are reported here in **IV. Observations**. Anomalous hypothetical symbol-based matched filter responses have been observed while a 26 foot diameter radio telescope beam transits the celestial direction of interest, 5.25 ± 0.15 hr Right Ascension (*RA*) and -7.6° ± 1° Declination (*DEC*), during 143 days of observations.

An important objective of this investigation is to automate the receiver's signal processing processes as much as reasonable. In past work, manual activities involved spreadsheet examination, sorting, and results plotting. In the current system, receiver signal file examination, filtering, sorting, statistical analysis and plotting of results are performed within the software system. Results reported here are the output of the multi-step signal processing machine, without human intervention, other than steps believed to be traceable to the machines' source code files, output files and lab notebooks.

A related objective of this investigation is to design machines that permit testing of various hyperparameters of the receiver system, for enhanced Bayesian data analysis, non-ergodic analysis, and Markov Chain model development, topics of intended further work.

Hypotheses and models, other than the RFI-augmented AWGN model, have not been introduced to an extent needed to try to explain the $\Delta t\ \Delta f$ anomalies. A lack of experimental evidence to support the RFI-augmented AWGN model leads to a belief that the cause of the repeated $\Delta t\ \Delta f$ anomalies are unexplained. Equipment, RFI, experimental methods, various types of bias and other explanations are being examined to attempt to explain the cause of the anomalies.

Past and present work emphasizes that many alternate and auxiliary hypotheses have not been modeled, tested and examined, to try to explain observed anomalies. Independent

---

William J. (Skip) Crilly Jr. is a Volunteer Science Ambassador in Education & Public Outreach
of the Green Bank Observatory. email: wcrilly@nrao.edu

corroboration of these experimental results is absent. Therefore, no conclusions can be made regarding the presence of an interstellar communication signal in these experimental results.

The following sections of this paper describe the hypothesis, experimental methods, observations, discussion, conclusions and further work.

## II. HYPOTHESIS

The hypotheses in previous work, [3][4], have been modified, as follows, to include an augmented AWGN explanation of anomalies in hypothetical discovery elements, including transmitter symbol repetition.

*Hypothesis*: An RFI-augmented AWGN model is expected to explain observations of one or more signal discovery elements contained within $\Delta t$ $\Delta f$ opposite circular-polarized pulse pairs, while a radio telescope beam diurnally scans the 0 to 6.3 hr Right Ascension range, including a direction of interest at celestial coordinates $RA$ $5.25 \pm 0.15$ hr and $DEC$ $-7.6° \pm 1°$.

The current hypothesis implicitly contains a choice of matched filter to be used at the receiver. The matched filter is defined to be the bandwidth and time duration of the energy detection system of the receiver, resulting from the FFT-based signal processing system, together with a filter that limits the range of polarized pulse pair $\Delta t$ and $\Delta f$ values, applied to statistical analysis, to seek symbol repetition in a population of quantized $RA$ directions. A symbol is defined in the context of this work to be a unique element among a set of possible pulse pair signals, indexed by quantized values of composite signal dimensions, including amplitude, RF frequency, pulse pair $\Delta t$ and $\Delta f$ values, pulse energy detection bandwidth and integration time, the latter two having fixed values in this work. Symbol repetition is defined to be the observed presence of one value of at least one hypothetically quantized dimension, repeated over time.

In past work, the matched filter used a 3.7 Hz binned FFT, 0.27 s integration, $|\Delta t| \leq 3$ s and $|\Delta f| \leq 400$ Hz, with a $|\Delta f| < 80$ Hz rejection region, the latter to ameliorate potential Doppler-spread RFI.[4] In the current work, the same matched filter is used to report follow-up measurements. In addition, two $\Delta t$ values, $\Delta t = -3.75$ s and $-6.25$ s, are used in the matched filter, together with other filter parameters set as above.

The stated uncertainty range of the $DEC$ of the hypothetical direction of interest, $\pm 1°$, is the value of the estimated uncertainty range in previous presentations, [3][4]. The radio telescope used in the current work has a measured power Full Width Half Maximum (FWHM) beamwidth, in the $RA$ plane, of approximately 2°, based on telescope continuum measurements made during Sun transits in the Fall of 2021. These calibrations had azimuth and elevation of the telescope set to values used during the beam transit tests described in this work. Circular beam symmetry is assumed. Therefore, the $\pm 1°$ $DEC$ uncertainty only applies when examining the prior experimental results, or current results analyzed under assumptions of significant prior results. The $RA$ uncertainty is based on the observation of anomalies in prior experiments.

Discovery processes in interstellar wideband communication are hypothesized to produce statistically anomalous changes in signal measurements, over time, i.e. differing from noise model and highly coded transmitter predictions. This changing signal concept is rationalized by considering an alternative argument, i.e. that the intended discoverable transmitter does not change its wideband signal properties. A continuously transmitting set of identical modulation properties, appearing to continuously have the properties of random noise, might communicate much information, yet does not lead a receiver designer to glean a method of decoding the signals to transmitted information. Therefore, it is expected that some discovery mechanisms, e.g. symbol repetition and matched filter changes, will be present in interstellar discoverable transmitter signals. In general, highly ergodic, low duty cycle signals present discovery difficulties. Some changes in signal properties, over time, are required, within the ensemble of transmission events, to aid in signal discovery and information recovery.

Changes in receiver settings during a multi-stage experiment tend to confound statistical analysis and Bayesian data analysis, because priors are potentially invalid during Bayesian posterior analysis. In addition, near-zero valued priors are problematic as they result in near-zero valued posteriors.[5] Hypothesis testing might yield no new information to guide further investigation, other than continued refutation of a prior explanatory model.

Standalone new observations, using new receiver settings, might yield highly unlikely noise-cause results, even when many possible changed receiver filters are factored in. For example, if a receiver's pulse pair $\Delta f$ search range is doubled from its prior experimental setting, and new anomalies are found, an RFI-augmented AWGN model likelihood may be assumed to increase by a factor of $N = 2$, to account for the two $\Delta f$ search range filter possibilities. In another example, if a single value of seemingly arbitrary pulse pair $\Delta t$ presents anomalies, and an $N$ quantity of $\Delta t$ values could have been chosen to seek anomalies, the factor $N$ may be used to increase the RFI-augmented AWGN model likelihood.

If a single $\Delta t$ value is found to have anomalous presence, and has other measured significance, e.g. anomalous repetition at highest values of SNR, within a single $RA$ range, then the single $\Delta t$ value may be examined for additional repetition, at lower SNR values. The use of one value of $N$, or another, depends on how arbitrary is the choice of one value of $\Delta t$ in a symbol repetition detection receiver.

In addition to a speculated discovery intent, interstellar communication signals are expected to contain some change in transmitted signal properties, to help the receiver further analyze the signal. An example of such hypothesized changes is a change in the matched filter required for reception. For example, a receiver might observe a disappearance of received signals, speculated to be due to a transmitter filter having changed. The receiver will presumably expect some significance to this event, and attempt to learn more about the transmitter signals, while attempting to understand why, and in what way, the transmitter filter appeared to change. The observed change might lead to understanding the information in the signal and the transmitter characteristics. In addition, changes in transmission properties might lead to unique identification of transmitters, when multiple transmitters may be present.

### III. METHOD OF MEASUREMENT

The receiver processes used in the current work adhere to the processes described in prior work, [3][4], with several changes, described in this section.

**Central IF rejection region increased**

Sporadic apparent spurious signals were observed in the region surrounding the 1425 MHz local oscillator. Leakage of low RF frequency signals into the IQ digitizers' inputs is suspected. The excised region surrounding the local oscillator is set to 1422 to 1428 MHz to prevent sporadic signals from affecting pulse pair AWGN likelihood calculations. The possible presence of associated spurious signals outside the excised range is an important aspect of equipment-cause hypotheses, requiring further study.

**RFI burst rejection filter IIR filter threshold increased**

The IIR narrowband RFI burst rejection filter, RFI amelioration method 3, in [3][4], was observed in the current work to reject single pulse pairs at high levels of SNR. The purpose of the filter is to reject rarely occurring repeated pulses, confined to a narrow range of RF bandwidth. The loop gain of the first-order IIR filter is therefore set to a high value, 0.99. A shortcoming of the filter is that a single pulse having a high SNR value can result in the filter output exceeding the $SNR_{IIR}$ threshold, without time integration, and a rejection of the pulse pair. Rejecting single pulses in pulse pairs is not desirable, because spectral outliers, rarely occurring $\Delta t$ $\Delta f$ polarized pulse pairs, are sought by the receiver. $SNR_{IIR}$ threshold was increased from 11.84 to 11.88 dB to ameliorate this effect.

**An energy burst filter is added**

In this work, wideband, short-time energy bursts have been observed in apparently random directions, MJD days, RF frequency and, in some $\Delta t$, $\Delta f$ measured values. The observed bursts appear similar to a group of pulse pairs observed on MJD 59332, described in Figures 9 and 10 in [4]. The hypothetical random energy burst observed pulse pairs in this work do not appear similar to observed standalone $\Delta t$ $\Delta f$ polarized pulse pairs. Rather, the observed energy bursts present multiple values of $\Delta t$ and $\Delta f$ in the polarized pulse pairs, and confined to a duration of several to multiple tens of seconds, without repeating in RA direction, and not appearing on subsequent days in the same RA direction. The energy burst pulse pairs appear to present $\Delta t$ values concentrated approximately within ± 1 s. Approximately ten discrete energy bursts were observed during the 143 days of the current experiment, at 6.3 hours of observation per MJD day. The cause of these receiver responses is unknown. The SNR measurement and thresholding method in the pulse detection system of the receiver uses different integration times and RF bandwidth for signal and noise measurements, [3][4] and is therefore sensitive to the high energy spectral outliers of energy bursts, resulting in associated observed $\Delta t$ $\Delta f$ pulse pairs in the receiver output.

A decision was made to design and implement a filter to separately measure high counts of $\Delta t$ $\Delta f$ responses observed on one day, occurring within a short time, to differentiate sporadic bursts of pulse pairs from apparent anomalous standalone $\Delta t$ $\Delta f$ polarized pulse pairs. The reasoning behind this decision is: 1) short duration wideband transient energy bursts are common phenomena in radio astronomy, and may be included in filters that augment the AWGN model, 2) transient receiver gain fluctuation, RFI and other mechanisms might cause the response, 3) the absence of observed repetition in the same RA direction, during multiple days, leads to an apparent absence of persistent interstellar communication signals, and 4) if an interstellar communication signal has properties that are similar to natural object signals, interstellar communication signal discovery is speculated to be difficult.

An SNR burst metric was developed to measure the concentration of burst-caused pulse pairs, binned by MJD, and by RA sub bins, each spanning 0.03 hr RA, resulting in 210 RA sub bins per MJD day, as follows,

$$SNR_{\text{BURST METRIC}}[MJD, RA_{subbin}]$$
$$\triangleq \sum \frac{(SNR_{\text{LHC}} + SNR_{\text{RHC}})}{2}, \quad (1)$$

where the summation includes $\Delta t$ $\Delta f$ pulse pairs present with less than 30 seconds of time to a subsequent pulse pair time, within an MJD and RA sub bin. $SNR_{LHC}$ and $SNR_{RHC}$ have decibel units, and have a minimum of 11.8 dB. LHC and RHC indicate left and right hand circular polarizations.

The energy burst rejection filter is set by performing a single set of measurements of the SNR burst metric on all MJD days, at pulse pair $|\Delta t| \leq 3$ s and $80$ Hz $\leq |\Delta f| \leq 400$ Hz. Subsequent pulse pair counts, using different $\Delta t$ $\Delta f$ filters, and/or trial count number, use the single set of SNR burst metric measurements, and a burst SNR metric threshold, set to 150, to reject pulse pairs observed within the indexed MJD and RA sub bin..

**Polarized pulse pair symbol repetition**

Signal processing methods filter candidate polarized pulse pairs into one of several possible repetition classes, given the number of degrees of freedom in the polarized pulse pair signals. The primary repetition class implemented in this work utilizes two $\Delta t$ values, and a range of $\Delta f$ values specified at $|\Delta f| \leq 400$ Hz, with a $|\Delta f| < 80$ Hz rejection region. The $\Delta f$ range is chosen to equal the $\Delta f$ range used in the 164 day and 40 day beam transit tests, and the 44 day artificial sky noise test previously reported [3][4]. $\Delta t$ values are indexed at 0.25 second quantization increments. $\Delta f$ values are generally quantized to the 3.725 Hz FFT bin width, with measured residuals less than one bin width. The residuals are due to quantization, metrology and calibration. RF frequency Doppler correction is referenced to 180 degree pointing azimuth at the Green Bank Observatory.

The values $\Delta t = -3.75$ s and $\Delta t = -6.25$ s are chosen for the repetitive symbol parameter. These two values were determined after observing an anomalously high number of high SNR repetitions in the hypothetical direction of interest at *RA* 5.1 to 5.4 hr, *DEC* -7.6°. The Δ*t* value is negative due to a chosen convention that Δ*t* measures the left-hand circular polarized pulse arrival time minus the right-hand circular polarized pulse arrival time, quantized to 0.25 s increments.

In the presentation of results, signal repetition measurements are plotted over a range of *RA* values from 0 to 6.3 hr, in 21 *RA* bins, or across a range of Δ*t* values in the *RA* direction of interest, observed during the 143 days.

**Energy burst repetition ameliorates transmitter-receiver filter mismatch**

Measured repetition of nearly equal values of Δ*t* is a method that differing receivers may use to discover repetitive signals in noise, without prior transmitter-receiver coordination. Identical repetitively transmitted signals result in identical repetitive observed spectral properties, at the output of an unchanging filter receiver, given a constant observation time that spans at least the duration of the energy burst, a fixed instantaneous bandwidth receiver, and the absence of noise and propagation channel fluctuations. When repetitive signals are present, two receivers, having different filtering processes, will each observe a repetition of outlying spectral properties, albeit different across receivers. In other words, a repetition receiver does not need to have a filter matched to a transmit filter, in order to detect a repetitive energy burst, transmitted in the ICH.

When noise is included in the repetitive received signal, and transmitter duty cycle considered, spectral outliers are expected to be observed using each of two differing receivers, with randomness estimated by Ricean statistics. Transmitted signals may have energy confined to narrow RF bandwidths to increase the number of pulses having low values of Δ*f*, with bursts transmitted at time spacing Δ*t*.

Using a repetitive transmit method, the issue of transmit and receive filter mismatch can be largely avoided, at the cost of energy required to repeat identical energy bursts. The cost may be justified if the repetitive signals are designed to contain information, using degrees of freedom in the signal properties.

**$SNR_{HIGH}$ vs. $SNR_{LOW}$ sorting has been replaced with SNR sorting using a new metric**

In current and past work, a question has arisen as to the choice of which polarization's SNR to sort, for the computation of likelihood functions. Sorting by $SNR_{HIGH}$, the higher of the two polarizations, has been used to measure likelihood values when searching for Δ*t* = 0 s polarized pulse pairs, while $SNR_{LOW}$ sorting has been used when searching within a range of Δ*t* ≠ 0 s values. The rationale for the two methods is explained in past work [4]. The two choices appear to not exhibit a potential sorting that optimizes a search for combined polarization high SNR outliers, given an AWGN model. In each method, SNR information about the pulses in pulse pairs is being discarded.

In the current work, a sorting metric has been produced that combines the left hand and right hand polarized SNRs into a single value, derived from the AWGN model likelihood of Rayleigh-distributed joint polarization signal amplitudes.

The sorting metric is designed while considering that the cumulative probability of power in Rayleigh distributed amplitude signals, above a given power value, is exponentially distributed as a function of the given power value. Polarization independence is assumed, creating a circularly symmetric joint distribution. A base ten logarithm function is added for convenience in relating the metric to orders of magnitude of decreasing likelihood,

$$SNR_{METRIC} \triangleq -\sum \log_{10} e^{-SNR_p} , \qquad (2)$$

where the summation is over *p*, the left-hand and right-hand circular polarization indexes. $SNR_p$ uses linear signal power to linear noise power ratio, after conversion from $(S_p+N_p)/N_p$ decibel measured values.

The SNR metric in (2) is based on the joint polarized pulses' cumulative likelihood of polarized pulse pairs observed at values equal to or higher than $SNR_{METRIC}$, given an AWGN model, Rayleigh amplitude distributed, and exponential cumulative power distributed. The use of the metric in sorting ameliorates the issue of potentially significant signal information lost when $SNR_{HIGH}$ and $SNR_{LOW}$ are separately used when sorting polarized pulse pairs during statistical analysis.

Rationale for the new metric considers the joint probability distribution of polarized RF tones combined with noise. **Figure 1** plots a calculated log likelihood function, based on the joint probability distribution function of the opposite polarized pair SNR levels, expected in an AWGN plus tone model, at three tone levels, 0, 1.5σ and 3.0σ, while σ is the standard deviation of the noise, normalized to one. The underlying mathematics is explained in Appendix B of [3]. The rightmost section of the plot favors the discovery of tone-carrying polarized pulse pairs.

The metric defined in (2) is expected to prioritize tone-carrying candidate pulse pairs observed most often at the highest sorted values of the metric, given the polarized signals' tone amplitudes. The pulse pair transmitted duty cycle is an implicit factor.

**A binomial probability distribution function is used to measure *RA* density likelihood**

Binomial probability calculations are used to estimate the likelihood that one *RA* range will contain a number of polarized pulse pairs, within a number of highest to lowest $SNR_{METRIC}$ sorted trials, including all measured *RA* ranges. The number of trials is equal to the rank, after the sort, of the measured $SNR_{METRIC}$ in the one *RA* range. The count of events seen is the number of filtered pulse pairs observed in an *RA* range, up to the trial number. The event probability used is one divided by the number of *RA* bins, equal to 1 / 21. Binomial density values that indicate a lower number of counts seen, in the trials, than expected due to noise, are set to equal the probability density value expected due to noise. Presentation of results plots therefore indicate anomalous presence of polarized pulse pairs, not their absence.

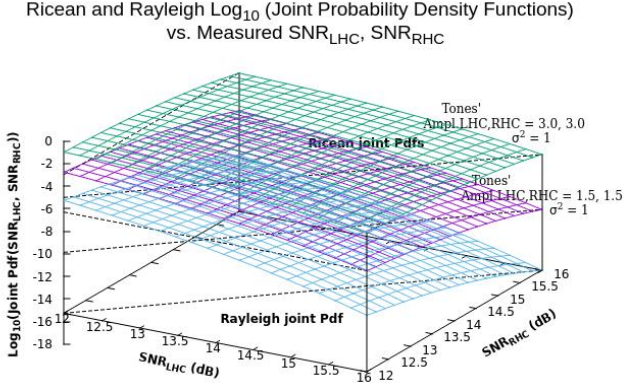

**Figure 1:** Calculated theoretical log joint probability density function values vs. $SNR_{LHC}$, $SNR_{RHC}$, where $SNR_{(\cdot)}$ is signal and noise power divided by noise power, measured at the output of each polarization receiver, at three levels of tone amplitudes, 0, 1.5 and 3.0, and where noise variance $\sigma^2$ is normalized to a value of one. Highest joint SNR values result in orders of magnitude decreased likelihood of an AWGN cause explaining polarized pulse pair events measured at highest SNR values. A Ricean amplitude probability distribution, Equation (7) in [3], Appendix B, was used to develop the probability density functions in the figure, together with a conversion of the measured $SNR_{LHC}$ and $SNR_{RHC}$ decibel values into amplitudes, and a $\log_{10}$ function applied to each density function calculated results.

**Signal processing data flow**
Three software steps are implemented, described as follows. 1. and 2. are described in [3][4].

   **1. FFT-based receivers, RFI amelioration**

   Radio telescope signals are digitized, and high SNR pulse characteristics are stored in four hour duration files per polarization, and per one second of a three second time interval. GPS clocks trigger six two-channel baseband IQ digitizers at the start of three time phases during a three second interval. RFI amelioration is described in past work. [3][4]

   **2. $\Delta t$ and $\Delta f$ search algorithm, RFI amelioration**

   Software searches of the output files of 1. are performed, logging polarized pulse pairs' properties having $|\Delta t|$ less than 10 seconds and $|\Delta f|$ less than 2 kHz, producing an output file for each MJD. The primary purpose of this step is to reduce the overall signal processing time, e.g. when different $\Delta t$ and $\Delta f$ filter hyperparameters are set in the next step.

   **3. Application of $\Delta t$ and $\Delta f$ filters, RFI amelioration**

   The output files of 2. are processed to produce a presentation of results for a set of hyperparameters. This process includes $SNR_{METRIC}$ sorting, hyperparameter filters, Right Ascension binning, RFI burst rejection, statistical analysis, and presentation of results development. The image files of the plots presented in **IV. OBSERVATIONS** are produced in the latter machine processing step.

## IV. OBSERVATIONS

**Figure 2** presents a follow-up of the experiments performed in prior work, [3][4], using new radio telescope data from the recent 143 day experiment. The modified augmented AWGN model is used, with the same $\Delta t$ and $\Delta f$ filter ranges used in the prior work. The *RA* range 5.1 - 5.4 hr does not indicate an anomalous presence of polarized pulse pairs, compelling a search for anomalies using larger values of $|\Delta t|$.

**Figure 3** presents an unlikely apparent repetition of a single value of $\Delta t$ = -3.75 s in the *RA* direction of interest. The absence of points above a value of -1 log likelihood is due to the presence of five of the ten highest $SNR_{METRIC}$ values, having $\Delta t$ = -3.75 s, among all 21 *RA* ranges, presented in the 5.1 to 5.4 hr *RA* direction.

**Figure 4** indicates an anomalous presence of filtered $\Delta t$ = -3.75 s pulse pairs during MJD days 59505 – 59569. An expected average AWGN-caused pulse pair flux of 1.5 pulse pairs per *RA* bin occurs in 65 days of observation, assuming that the non-bin 17 points in **Figure 4** are explained by noise events. The presence of eight pulse pairs in the *RA* bin 17 direction in 65 days seems anomalous, having binomial likelihood 0.004. On the other hand, the eight pulse pairs might be explained as random population selection effects.

When the log likelihood of the sorted $SNR_{METRIC}$ pulse pairs is re-calculated for a sample population of MJD 59500 - 59569, the *RA* bin 17 log likelihood value decreases from the minimum value indicated in **Figure 3** to -5.29. The five highest $SNR_{METRIC}$ pulse pairs appear in the *RA* direction of interest, i.e. 5.1-5.4 hr *RA*.

**Figure 5** indicates that polarized pulse pairs are distributed somewhat uniformly across RF frequency. This behavior is expected due to an AWGN model, and due to a model of a high information capacity communication signal.

**Figure 6** presents $\Delta f$ measurements of polarized pulse pairs having the highest 70 values of $SNR_{METRIC}$ at $\Delta t$ = -3.75 s during the 143 day experiment, binned to 21 *RA* ranges. The highest eight $SNR_{METRIC}$ polarized pulse pairs in *RA* bin 17 (5.1-5.4 hr *RA*) appear to present a repetition of $\Delta f$ values at quantized multiples of ±58.575 Hz.

Monte Carlo simulations of uniform distributed $\Delta f$ values, within the $\Delta f$ filtered range, estimate a 58.575 Hz multiplier event probability of 0.259, and a standard deviation of residuals, $\sigma_{RESIDUALS}$, measured at 8.43 Hz. In other words, the probability of a single pulse pair having a multiplier event residual less than or equal to 8.43 Hz, given a 58.575 Hz multiplier, simulated to 0.259. A binomial likelihood function was used to estimate the event probability in the Monte Carlo method, using 1000 trials. The Monte Carlo derived event probability, 0.259, may be compared to an intuitive probability estimate of twice the $\sigma_{RESIDUALS}$ divided by 58.575 Hz. The 0.259 value calculates to 1.8 $\sigma_{RESIDUALS}$ / 58.575. The event probability estimated in this process is used, as follows, in a likelihood function estimating multiple event probabilities that each appear to fit a multiplier pattern.

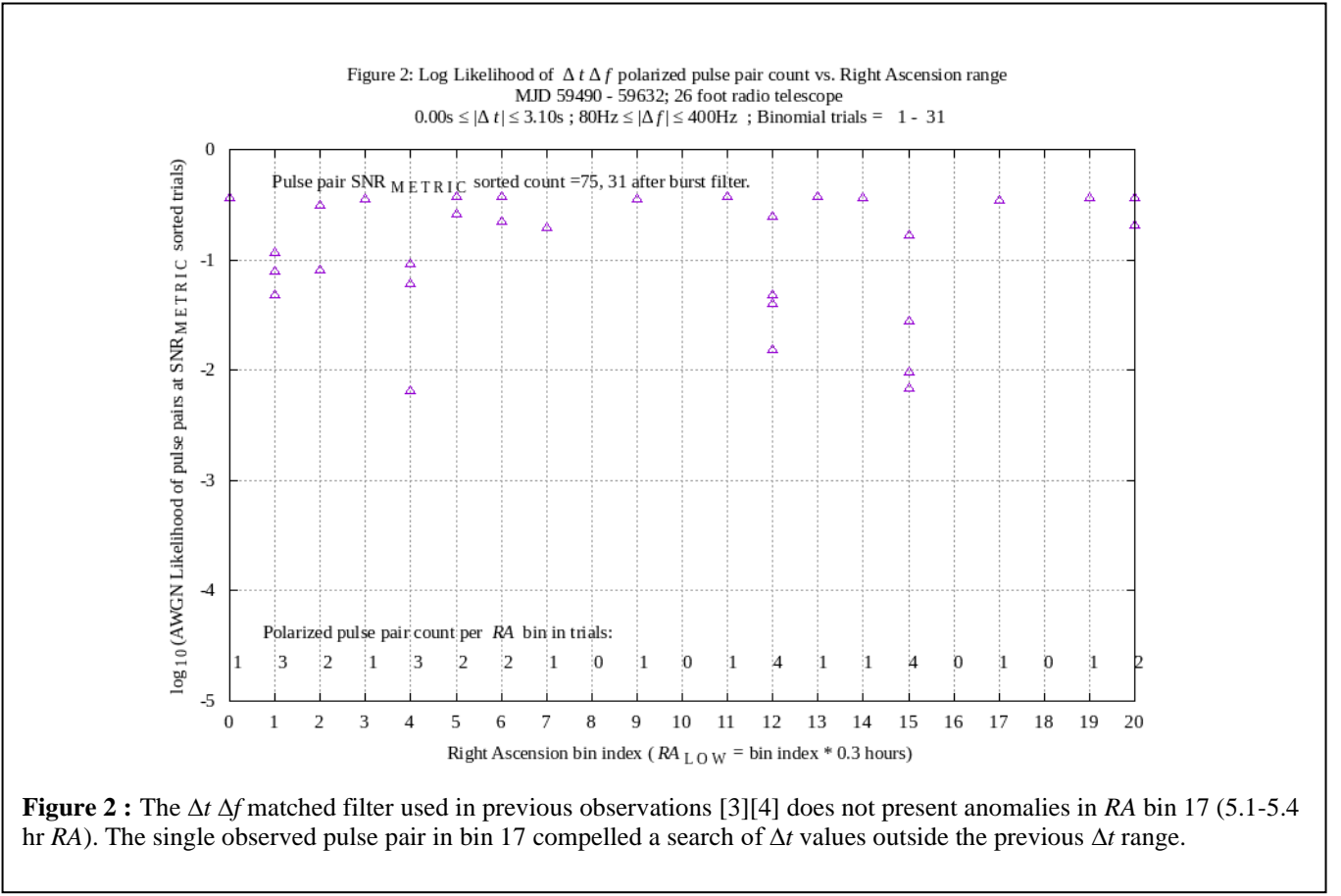

**Figure 2 :** The $\Delta t\ \Delta f$ matched filter used in previous observations [3][4] does not present anomalies in *RA* bin 17 (5.1-5.4 hr *RA*). The single observed pulse pair in bin 17 compelled a search of $\Delta t$ values outside the previous $\Delta t$ range.

The $\Delta f$ residuals of the eight highest $SNR_{METRIC}$ $\Delta t = -3.75$ s pulse pair events in the 5.1-5.4 hr *RA* direction measure within ± 10.25 Hz, using a ±58.575 Hz $\Delta f$ base multiple. Using the Monte Carlo above-derived estimate of event probability, at 1.8 $\sigma_{RESIDUALS}$ / 58.575, computed at $\sigma_{RESIDUALS}$ = 10.25 Hz, yields a pulse pair multiplier event probability of 0.315.

The binomial distribution of eight events seen in eight trials, at a probability of 0.315, calculates to $9.7 \times 10^{-5}$. Various factors significantly increase this value. For example, the freedom in the choice of base multiplier $\Delta f$ values may be estimated to be the ratio of 58.575 to the FFT bin width in the signal processing system, resulting in a factor of 58.575 Hz / 3.725 Hz = 16. The use of this factor in calculating likelihood increases the augmented AWGN model likelihood of the eight polarized pulse pairs presented in the experiment, to 0.0015. Another consideration is that some of the eight values are expected to be noise-caused, because one noise-caused pulse pair is expected on average, in a number of trials equal to the number of *RA* bins, 21. In this case, perhaps only five pulse pair events, ranked 2, 3, 6, 7 and 11 in $SNR_{METRIC}$, should be used when calculating likelihood. Five events seen in five trials at the event probability of 0.315 estimates the 58.575 quantized pulse pairs' likelihood present in the experiment at 0.0031, and 0.05 using sixteen base multiplier values.

The number of trials used for binomial calculations is the number of pulse pairs output from the energy burst filter. The numbers of these are posted near the top of each plot. Relative differences from plot to plot occur due to different $\Delta t$ filters used in plots.

**Figures 7 - 10** plot log likelihood, MJD, RF frequency and $\Delta f$ measurements of the apparent repeated $\Delta t = -6.25$ s polarized pulse pairs, observed in *RA* bin 17.

**Figure 11** plots values of $|\Delta t| \leq 1.1$ s, speculated to be present if a natural object produces bursts of energy into the radio telescope receiver. The direction of interest does not indicate anomalies. Models of natural objects and associated receiver response are preliminary.

**Figure 12** plots the MJD of a range of $\Delta t$ values from -8 to –3 s, containing the two anomalous $\Delta t$ values at -3.75 s and -6.25 s. A search of other $\Delta t$ values, to ± 10 s, did not result in an indication of anomalous pulse pairs in the *RA* direction of interest. The elements of two sets of pulse pairs, each set having either $\Delta t$ values of -3.75 s and -6.25 s, do not overlap with the same MJD, except for two pulse pairs at 59546.2542303 and 59546.2540972, respectively, occurring 11.25 s apart. In **Figure 5**, two $\Delta t = -3.75$ s pulse pairs are observed relatively close in RF frequency, 1403.3933225 MHz on MJD 59546 and 1403.1295031 MHz on MJD 59565. The former pulse pair is the pulse pair that indicates $\Delta t = -3.75$ s and $\Delta t = -6.25$ s overlap on 59546 MJD. The presence of the MJD overlap, close RF frequency spacing, close time spacing, leads to a conjecture that pulse pairs might be transmitted in bursts spanning a few hundred kHz, and having durations of a few tens of seconds.

Symbol repetition in interstellar communications: methods and observations

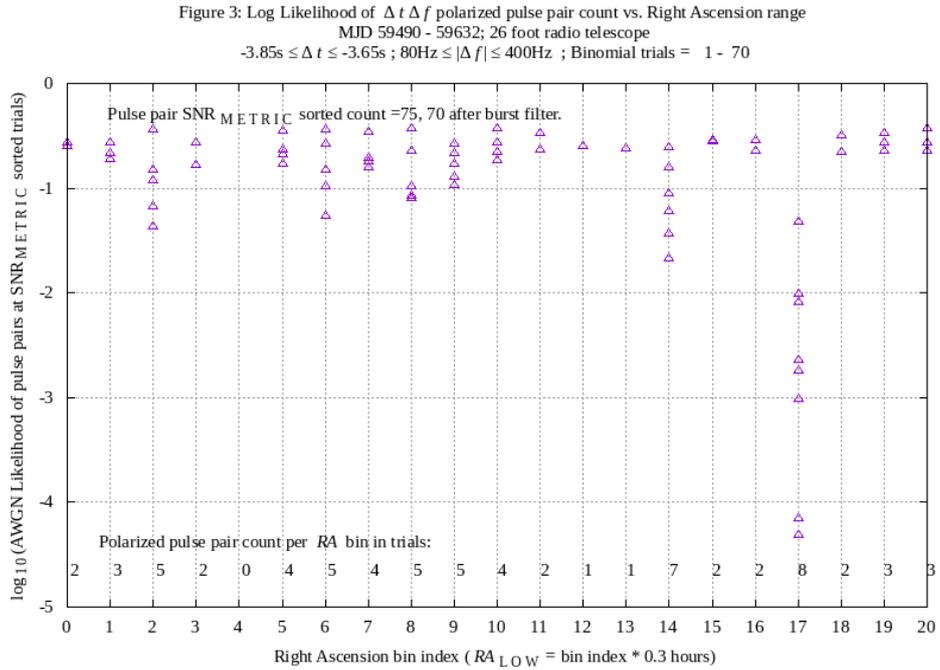

**Figure 3:** A $\Delta t$ = -3.75 s matched filter presents an anomalous number of highest SNR$_{METRIC}$ polarized pulse pairs. The *RA* 5.1 to 5.4 hr polarized pulse pairs ranked 1, 2, 5, 6, 10 in the SNR$_{METRIC}$ highest 10 values across all *RA*s. The AWGN likelihood of the direction of interest observations calculate approximately 1000 times less than expected.

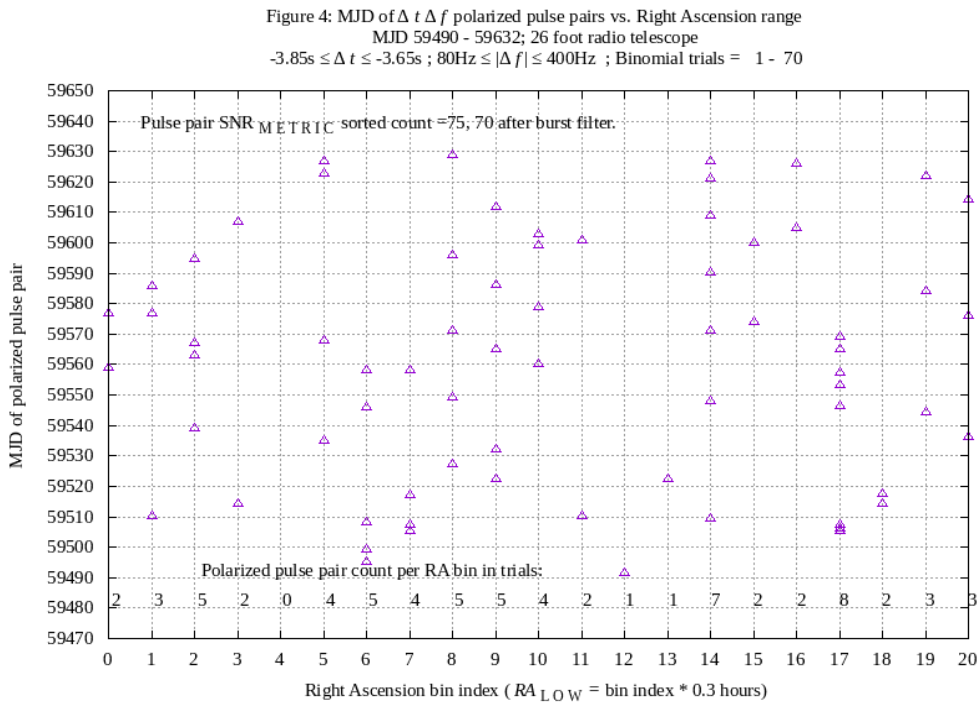

**Figure 4:** The 5.1 to 5.4 hr *RA* anomalous $\Delta t$ = -3.75 s polarized pulse pairs were observed on MJD days 59505, 59506, 59507, 59546, 59553, 59557, 59565, and 59569. The bin 17 MJD concentration appears unusual and non-ergodic**.**

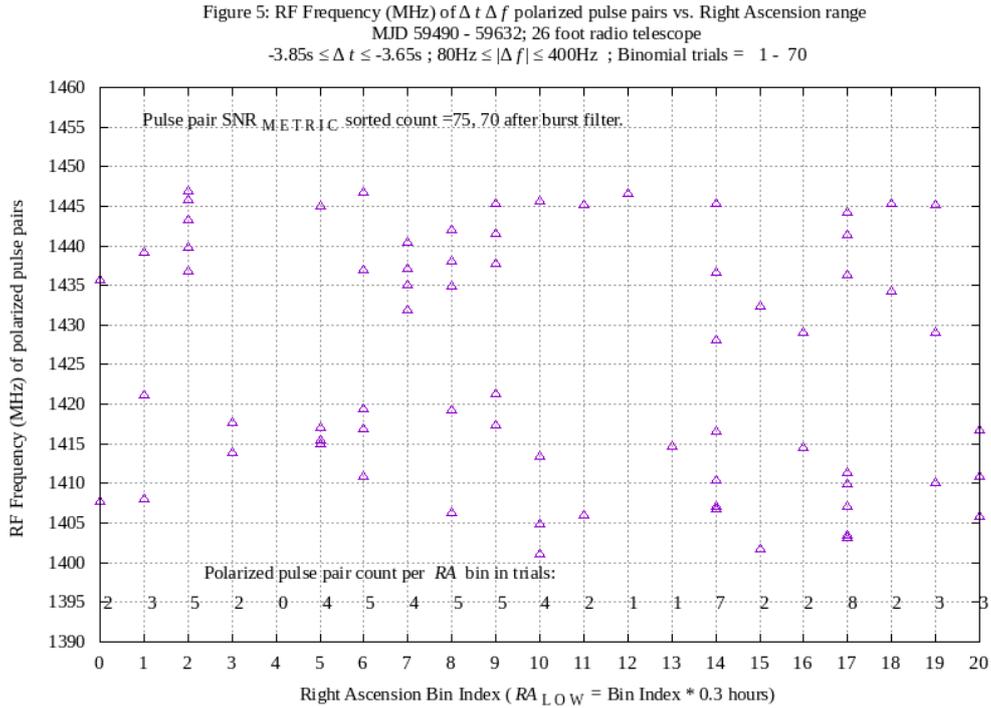

**Figure 5:** Measured RF frequency of the polarized pulse pairs appear distributed across the receiver range. A central region of 1422 to 1428 MHz was excised due to suspected RFI at low intermediate frequencies, surrounding the local oscillator. Two *RA* bin 17 points measured 1403.3933225 MHz on MJD 59546 and 1403.1295031 MHz on MJD 59565.

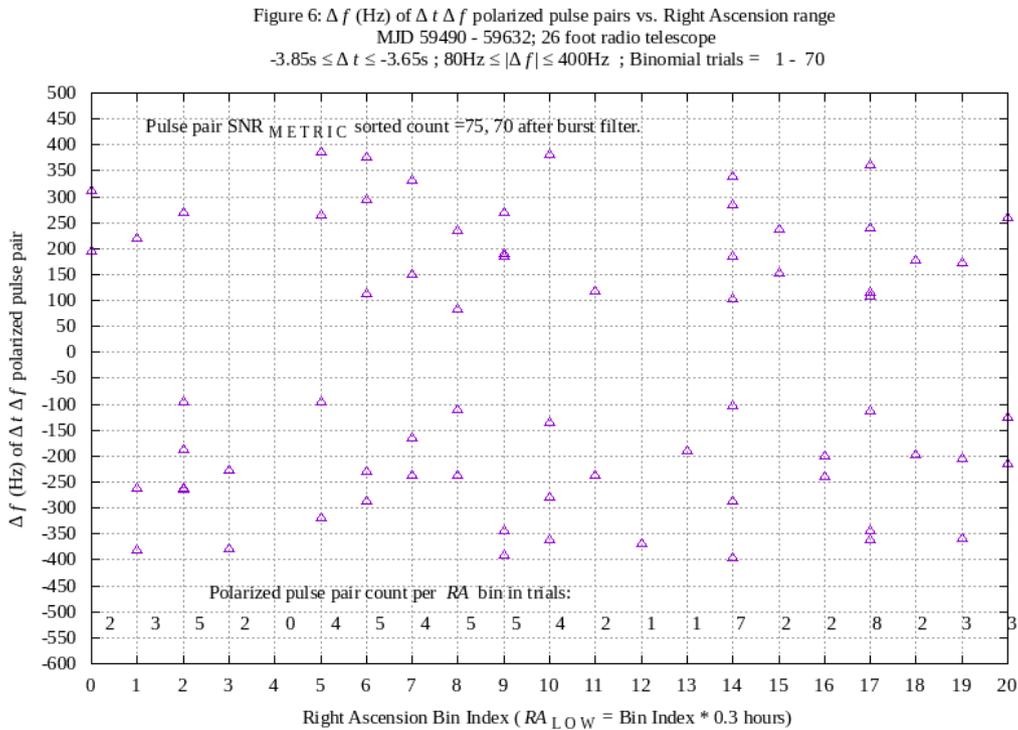

**Figure 6:** The eight anomalous $\Delta t = -3.75$ s pulse pair events appear concentrated near multiples of $\Delta f = 58.575$ Hz, and measure maximum absolute residuals $\leq 10.25$ Hz. The binomial distribution of 8 events seen in 8 tries, at event pr.0.315, calculates to $9.7 \times 10^{-5}$. Sixteen possible multiplier values increase this AWGN model likelihood to 0.0019.

Symbol repetition in interstellar communications: methods and observations

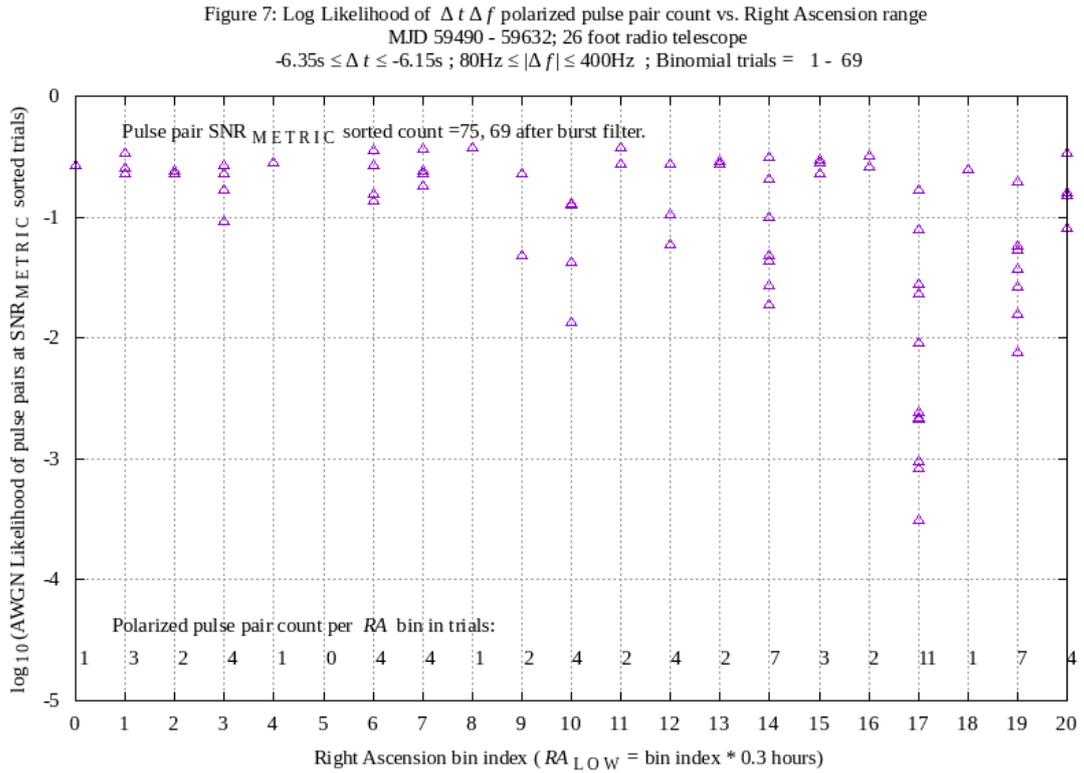

**Figure 7:** A $\Delta t$ = -6.25 s matched filter presents anomalous log likelihood measurements in *RA* bin 17.

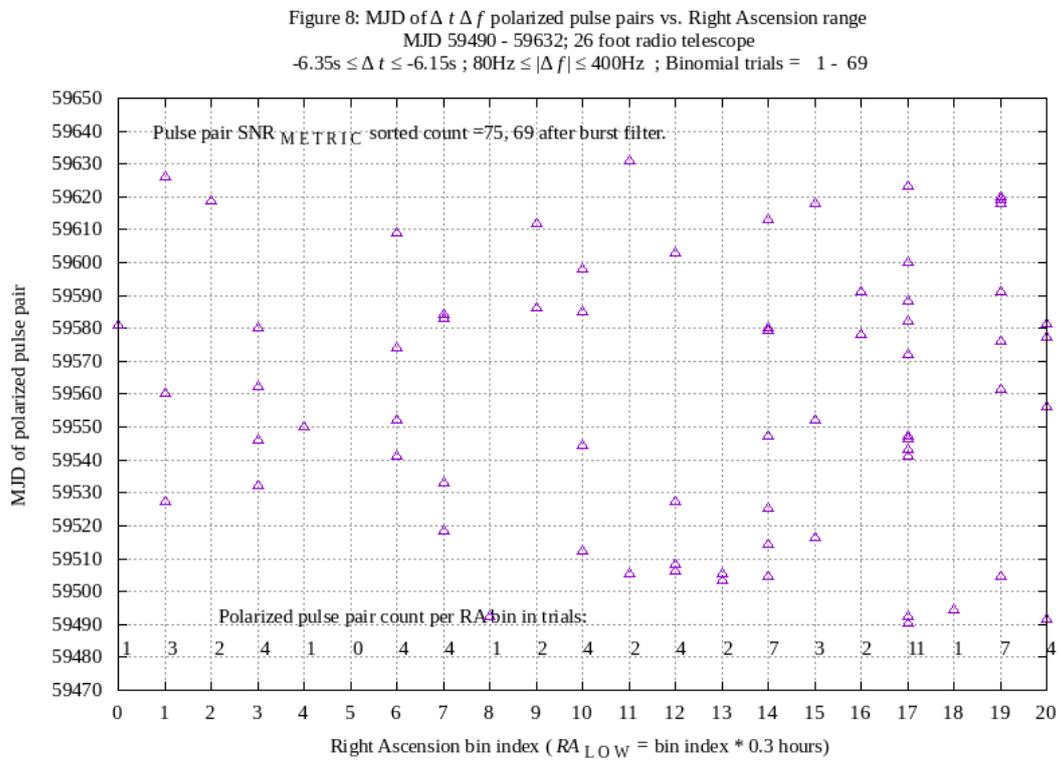

**Figure 8:** Polarized pulse pairs in *RA* bin 17 and $\Delta t$ = -6.25 s, appear concentrated in MJD ranges having almost no overlap of the MJD ranges observed with $\Delta t$ = -3.75 s polarized pulse pairs. **Figure 12** describes these relative MJD concentrations.

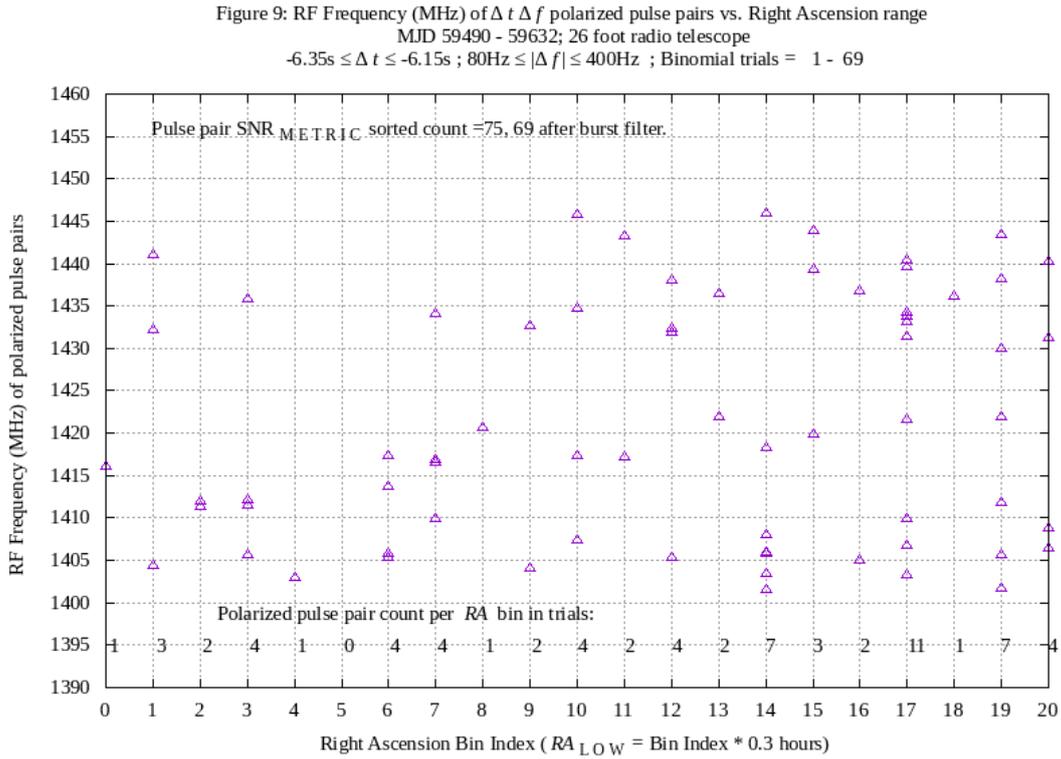

**Figure 9:** $\Delta t$ = -6.25 s polarized pulse pairs appear to be largely distributed across RF frequencies, while in the direction of interest, three pulse pairs were observed at 1433.1568978, 1433.8466790, and 1434.2319196 MHz.

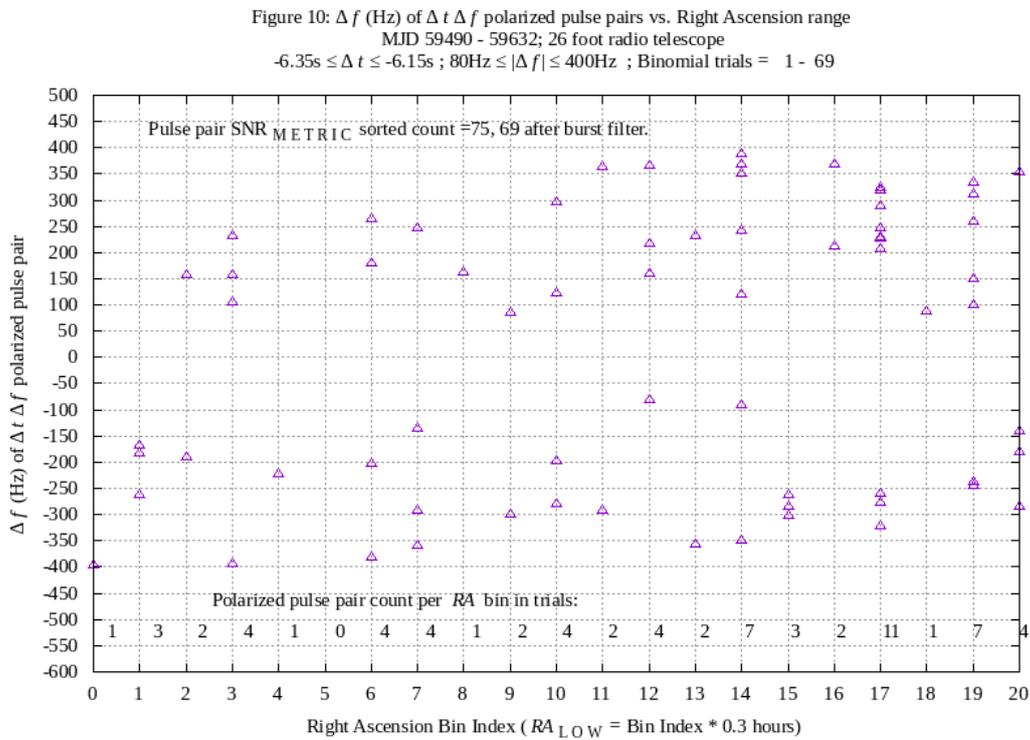

**Figure 10:** Eleven $\Delta t$ = - 6.25 s polarized pulse pairs observed in the 5.1 – 5.4 hr $RA$ region appear to have $|\Delta f|$ concentrated between 200 - 350 Hz, approximately half the hyperparameter filter range. This $|\Delta f|$ concentration may be estimated to occur with a probability of one in $2^{11}$ tries, multiplied by two, to account for two possible hyperparameter ranges. Three of the eleven $\Delta f$ values appear to repeat close in value.

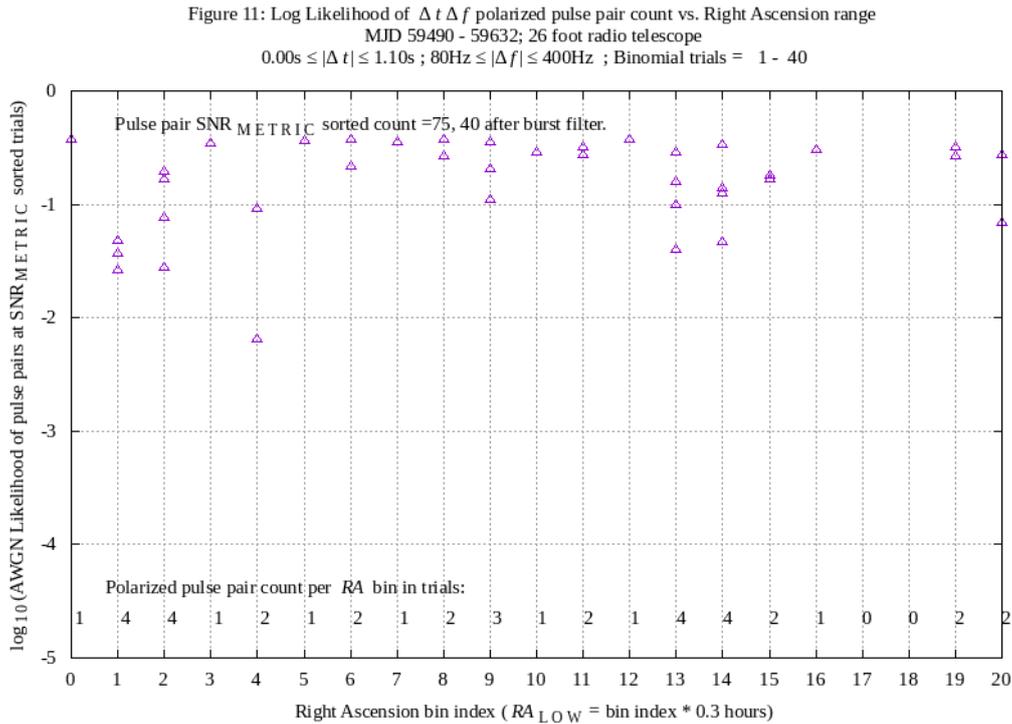

**Figure 11:** Assuming the $\Delta t = -3.75$ s and $\Delta t = -6.25$ s repetitions are explained by one or more unpolarized repetitive natural objects, pulse pairs would be expected to be observed in at least a few of the nine 0.25 s quantized $\Delta t$ values having $|\Delta t| \leq 1.1$ s, in *RA* bin 17. Such responses are absent in the measurement of the direction of interest, indicating that spectral power outliers, due to a natural source, are not near-simultaneously present in opposite circular polarizations.

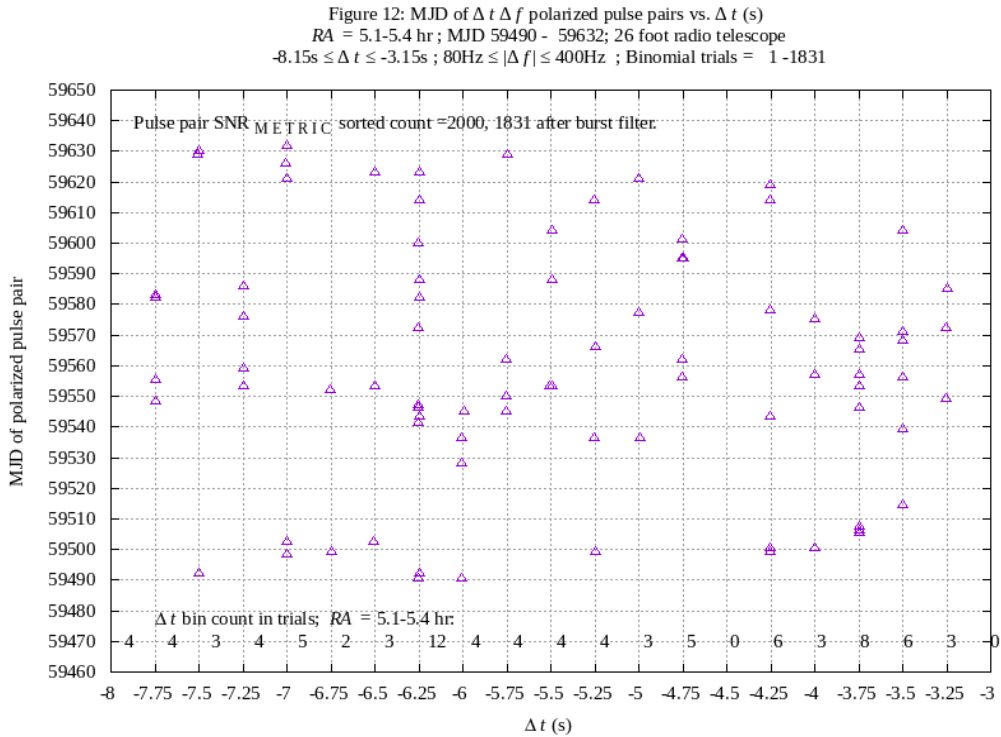

**Figure 12:** The $\Delta t = -3.75$ s and $\Delta t = -6.25$ s apparent symbol repetitions are concentrated in non-overlapping MJD groups, with one exception, at MJD 59546.2542303 and 59546.2540972, respectively, 11.25 s apart. The $\Delta t = -3.75$ s pulse pair is one of the two pulse pairs having close RF frequencies, described in text in **Figure 5**.

## V. Discussion

There are a number of issues that potentially confound the analysis of signals and anomalies identified in this work.

**Anomaly identification potentially has observer bias**

Some of the anomalies mentioned in this work have associated stochastic calculations, while others do not. Confirming the true nature of an anomaly requires one to consider many possible explanations, and combinations of explanations.

**Population selection bias**

The selection of $\Delta t$ candidate values, based on concentration of $\Delta t$ data, leads to a concern of sample collection bias. For example, $\Delta t$ is quantized to 0.25 second steps, and the $\Delta t = -3.75$ s value is three quantized values outside the $\Delta t$ range used in the prior experiment [4]. Considering positive and negative possible values, a factor of six needs to be applied to measured likelihood values, calculated using a noise-cause model.

**Leakage of RFI and burst filters**

The RFI and natural object burst filters may be allowing anomalous responses to indicate. Measurement data and output files produced by these filters need to be examined in a systematic way to develop metrics that seek to refute this auxiliary hypothesis that might explain anomalies.

**Natural objects require models**

Repeating indications at values of Right Ascension implies a celestial source. A celestial hypothesis may be tested using simulations and measurements. This work is required before natural object likelihood can be quantified.

**Geostationary satellite RFI**

The radio telescope pointing direction is near the geostationary Clarke Belt. A satellite-common RFI cause seems unlikely to show *RA* concentration during 143 days, but might be present.

**Human error and human-made machine errors**

The benefit of all-machine processing raises the countervailing issue of human error made when writing source code, albeit the latter generally having better traceability and testability.

There are anomalies in this work that suggest a conjecture that the *RA* 5.1 - 5.4 hr pulse pairs might be caused by an intentional low duty cycle transmitter having up to a few hundred kHz of RF bandwidth, and a few tens of seconds of transmit duration, indicated by MJD, time and frequency range overlap of pulse pairs in observations shown in **Figures 5, 8 and 12,** and Green Bank and Haswell associated pulses reported in [3] Figures 7, 12 and 13. The anomalous $\Delta t = -6.25$ s , $|\Delta f|$ concentration of pulse pairs shown in **Figure 10** is difficult to explain, due to noise or natural objects, and further leads to conjectures about hypothetical transmitter signal characteristics and receiver equipment causes.

## VI. Conclusions

The AWGN model, augmented with extensive RFI amelioration, does not adequately explain the observed results. A large number of alternate and auxiliary hypotheses may be considered to potentially explain the anomalies seen in this experiment. Further work is required.

## VII. Further Work

1. Further work described previously. [3][4]
2. Markov Chain analysis, using Monte Carlo methods [5].
3. The apparent non-ergodic properties of signals compels a need to create signal source models and processes that may be used to test various transmitter and natural object scenarios against observed results.
4. Continue radio telescope receiver captures at -7.6° DEC, and continue signal analysis.
5. Search additional signal dimensions for symbol repetition. Check various hyperparameter ranges.
6. Perform tests to verify equipment and software operation.
7. Help others seeking replication and corroboration.

## VIII. Acknowledgements

Many people helped make this and previous work possible, and highly enjoyable. The author is grateful for the many contributions of workers of the Green Bank Observatory, Deep Space Exploration Society, Society of Amateur Radio Astronomers, SETI Institute, Berkeley SETI Research Center, Breakthrough Listen, product vendors, and open source software community. The author is grateful for guidance and encouragement from family and friends.